# Quasi-quantization of writhe in ideal knots

Piotr Pierański and Sylwester Przybył,

*Faculty of Technical Physics*
*Poznań University of Technology*
*Piotrowo 3, 60 965 Poznań,*
e-mail: Piotr.Pieranski@put.poznan.pl

**ABSTRACT**

The values of writhe of the most tight conformations, found by the SONO algorithm, of all alternating prime knots with up to 10 crossings are analysed. The distribution of the writhe values is shown to be concentrated around the equally spaced levels. The "writhe quantum" is shown to be close to the rational 4/7 value. The deviation of the writhe values from the n*(4/7) writhe levels scheme is analysed quantitatively.

## 1. Introduction

Knot tying is not only a conscious activity reserved for humans. In nature knots are often tied by chance. Thermal fluctuations may entangle a polymeric chain in such a manner that an open knot becomes tied on it. This possibility and its physical consequences were considered by de Gennes [1].

Whenever the ends of a polymer molecule become connected – a closed knot is tied. Understanding the topological aspects of the physics of polymers is an interesting and challenging problem [2, 3]. Formation of knotted polymeric molecules can be simulated numerically [4, 5]. A simple algorithm creating random walks in the 3D space provides a crude model of the polymeric chain. Obviously, knots tied on random walks are of various topological types. The more complex a knot is, the less frequently it occurs. The probability of formation of various knot types was studied [6] and the related problem of the size of knots tied on long polymeric chains was also analysed [7].

DNA molecules are not protected from becoming knotted. Knots of various types and catenanes are easily formed as intermediate products of DNA replication and recombination processes [8]. The probability of the knot formation within the DNA molecules was analysed [9]. In appropriate experimental conditions knots tied on DNA molecules of identical molecular weight can be created [10]. Although being of the same molecular weight, the topoisomers display different physical properties. For instance, their electrophoretic migration rate is different and proportional to the average crossing numbers of the so-called ideal conformations of the knotted molecules [11].

Tying a knot on a rope needs closing its ends. In the case of DNA, the closing is easier if meeting ends are appropriately oriented. The geometrical parameter of a knot responsible for the relative orientation of its ends is the writhe of the knot [12, 13, 14]. As shown recently, writhe of the ideal conformations of prime knots displays a curious quasi-quantization properties. Below, we present results of our study of the phenomenon.

## 2. Ideal knots

From the topological point of view a conformation of a knot is of no importance [15]. From the physical point of view it matters a lot. For knots tied on the ideal, i.e. the utterly flexible but infinitely hard in its circular cross-section rope there exists a minimum value of the rope length at which knot of the given type can still be tied. The particular conformation of a knot for which the minimum is reached is called *ideal* [16]. It is assumed that there exists but a single conformation which minimizes the rope length.

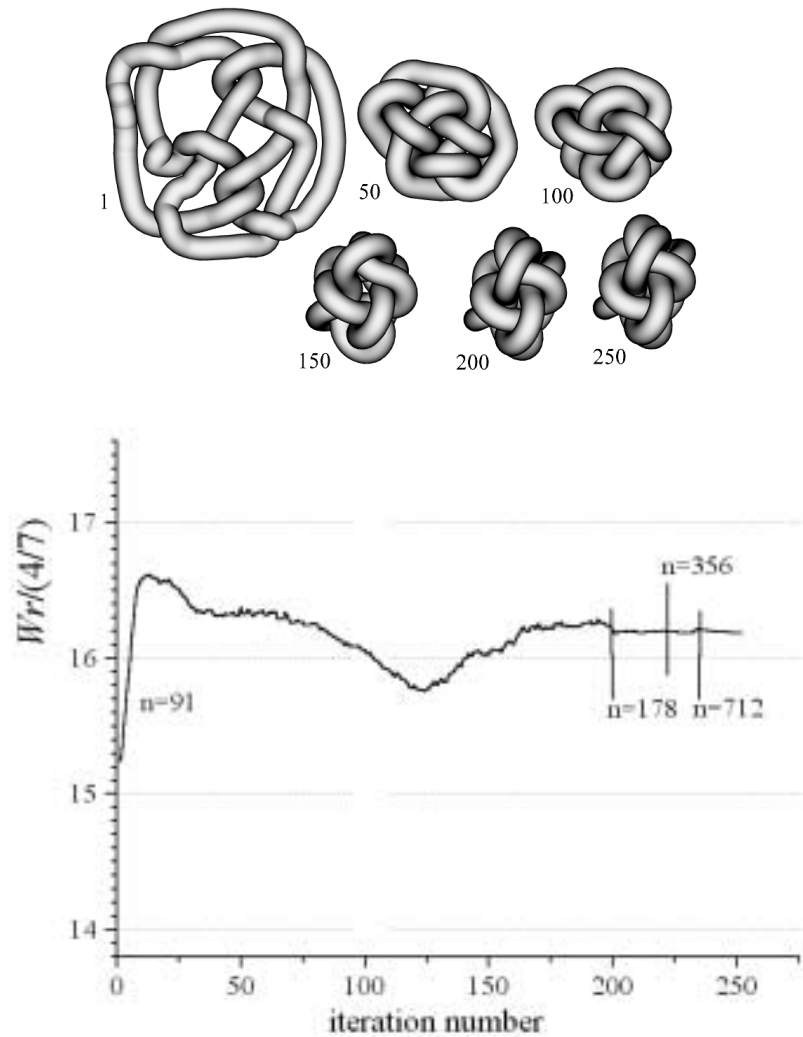

Fig. 1. Evolution of a loose conformation of the $10_{120}$ knot towards its most tight conformation. The evolution was enforced by SONO algorithm. Calculations started with a number of segments $N=91$. At the end of the tightening process the number of segments was doubled 3 times up to $N=712$. Numerical errors within the final value of writhe are smaller than 1% and the deviation of the writhe value from 16 predicted by the Cerf-Stasiak is an intrinsic property of the conformation found by SONO.

Finding ideal conformations is not a trivial task. There exist a few algorithms which are aimed to perform it. One of them is SONO (Shrink-On-No-Overlaps) described in our earlier papers [17]. Fig.1 presents how simulating the process in which the rope shrinks slowly SONO arrives at the ideal conformation of the $10_{120}$ knot. As seen in the figure, the knot changes considerably its conformation. The changes of conformation are accompanied by changes of its writhe value.

Initially, to speed up the evolution, the number of segments, of which the equilateral knot is constructed, is kept as low as possible. At the end of the calculation it is raised to a value at which the inaccuracy of the writhe calculation is better than 1%.

Although for the sake of brevity we talk in what follows about the ideal conformations, the "*ideal*" must be understood as "*the most tight one, found by the SONO algorithm*". There is no escape from this uncertainty - the ideal conformation is known at present only for a single prime knot: the trivial knot.

## 3. Writhe

One of the essential parameters which distinguish between the shapes of various conformations of the same knot is the 3D *writhe*. (In what follows we shall refer to it in brief as *writhe*.) If $\mathbf{r}_1$ and $\mathbf{r}_2$ are two points within conformation $K$ of a knot and $\mathbf{r}_{1,2} = \mathbf{r}_2 - \mathbf{r}_1$ is the vector which joins the points, then writhe of the conformation is given by the value of the double integral:

$$Wr = \frac{1}{4\pi} \oint_K \oint_K \frac{(d\mathbf{r}_1 \times \mathbf{r}_{1,2}) \cdot d\mathbf{r}_2}{r_{1,2}^3} \qquad (1)$$

As shown by Georges Călugăreanu, in spite that at the $\mathbf{r}_1 = \mathbf{r}_2$ diagonal of the $K \times K$ integration domain the denominator of the integrated ratio goes to zero, the integral does not diverge.

Writhe calculation formula defined by eq.1 is valid for continuous knots. Knots processed by the SONO algorithm are discrete, they are represented by tables of (*x,y,z*) coordinates of *n* vertices. Thus, the calculation of their writhe must be performed using discrete sum formulae [18].

## 4. Writhe quantization hypothesis

In the first paper on the geometry of ideal knots [19], it has been indicated that the writhe within some families torus and twist prime knots grows with the crossing number in a linear manner. Ideal conformations of knots discussed in ref. 19 were found via simulated annealing. Using the more efficient SONO algorithm we performed an extensive search for the most tight conformations of all prime knots with up to 10 crossings. Preliminary analysis of the results we obtained for knots with up to 9 crossings was described in ref. 17 where we pointed out (see fig. 8 there) that the writhe values of prime knots with up to 9 crossings show a visible tendency to gather around a few, well defined levels. No hypothesis concerning their spacing was formulated.

The observation stimulated a series of theoretical considerations [20], which lead to the hypothesis that for alternating prime knots the quantum of writhe is 4/7, and showed that the actual writhe values can be predicted by a topological invariant computable from any minimum crossing number diagram. It is the aim of the present paper to verify quantitatively the hypothesis of the 4/7 writhe quantum on the set of all alternating prime knots with up to 10 crossings.

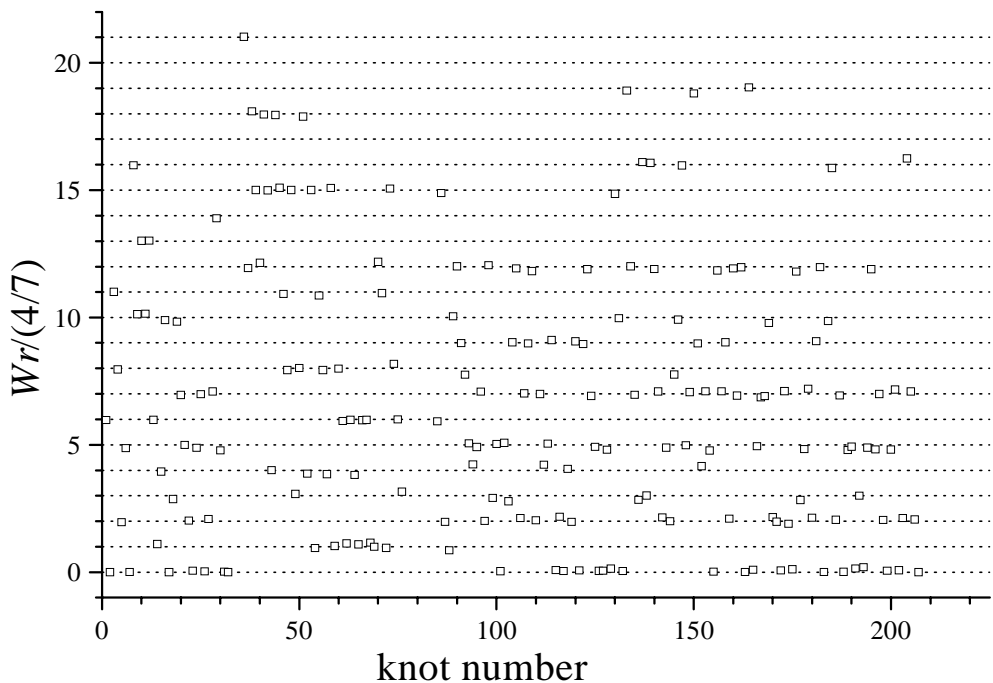

**Fig. 2.** $Wr/(4/7)$ *versus* the knot number for ideal conformations of all alternating prime knots with up to 10 crossings. Horizontal lines indicate the writhe levels suggested by Cerf and Stasiak. The gap visible in the set of plotted points in the vicinity of the knot number=80 corresponds to the non-alternating knots with 9 crossings. A smaller gap localized around knot number=40 corresponds to 3 non-alternating knots with 8 crossings.

## 5. Quantitative verification of the 4/7 hypothesis

If the Cerf-Stasiak 4/7 writhe quantum hypothesis is right, the values of $Wr/(4/7)$ should for the alternating knots be located close to integer levels. Fig. 2 presents the plot of the $Wr/(4/7)$ values of the ideal conformations of all alternating prime knots with up to 10 crossings. The values are plotted versus the *knot number* which localizes a particular knot in the Rolfsen table of prime knots. See, e.g. ref. 21. Confirming our earlier observations reported in ref. 3 the plot reveals that the writhe values are distributed in a highly inhomogeneous manner. The values of the $Wr/(4/7)$ variable are clearly gathering around the integer levels; the Cerf-Stasiak 4/7 writhe quantum hypothesis seems to be qualitatively confirmed. The $n*(4/7)$ writhe levels scheme was plotted in Fig. 2 according to the suggestions of Cerf and Stasiak. It seems to fit well the data provided by SONO. But, is it really the best writhe levels scheme? A simple test convinced us, that this is indeed the case.

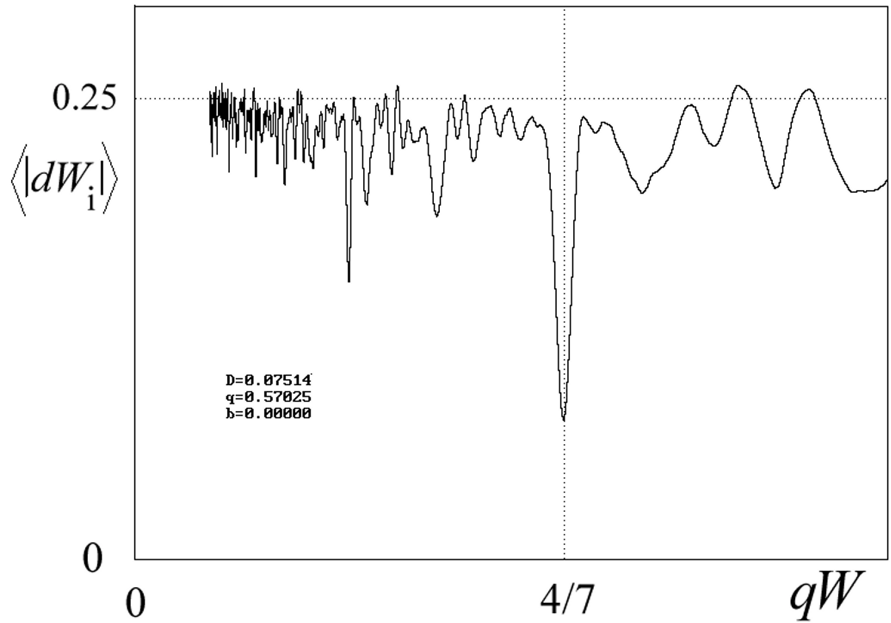

**Fig. 3.** The average relative deviation $\langle|dW|\rangle$ of the writhe values of all alternating knots with up to 10 crossings from $n*qW$ writhe levels *versus* the writhe quantum $qW$.

To check quantitatively which value of the writhe quantum $qW$ fits best the set of our writhe data, we calculated the dependence of the average relative deviation $\langle|dW_i|\rangle$ of the writhe values $Wr_i$ from $n*qW$ levels, where $qW$ was swept throughout the [0.1, 1.0] interval. The $\langle|dW|\rangle$ value was calculated as follows:

$$\langle|dW|\rangle = \frac{1}{N}\sum_{i \in A}|dW_i| = \frac{1}{N}\sum_{i \in A}\frac{|Wr_i - n_i^{best} qW|}{qW} \qquad (2)$$

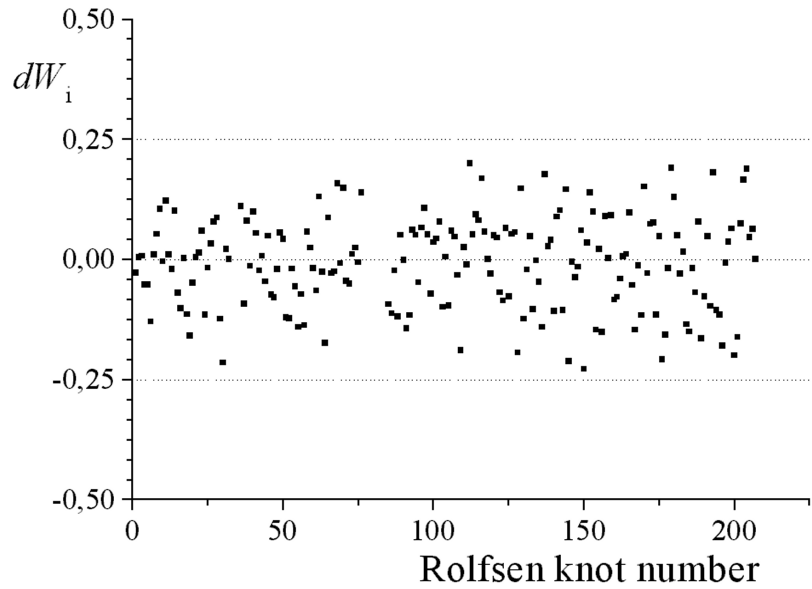

**Fig. 4.** The deviations $dW_i$ of the writhe values of all alternating knots with up to 10 crossings from levels of the optimal scheme defined by $qW=4/7$.

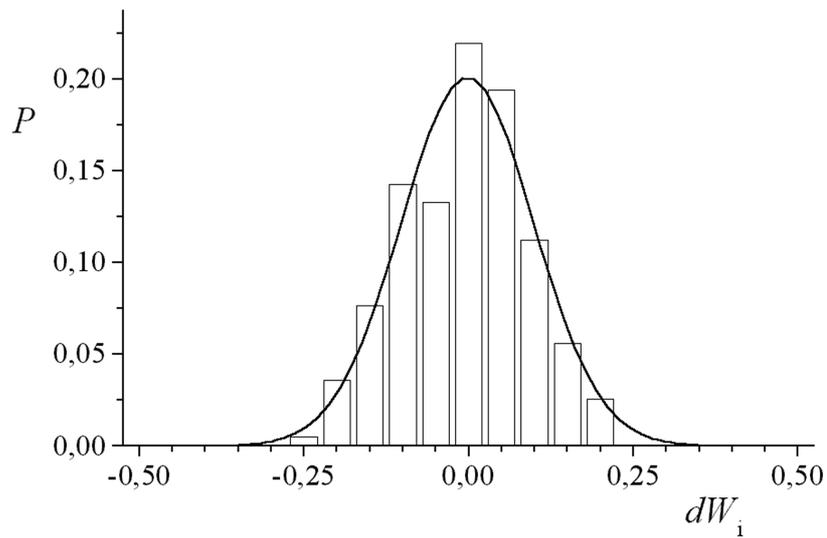

**Fig. 5.** The probability $P$ of finding a the $dW$ deviations within 20 counting bins. The Gaussian function which fits the data is also plotted; its half-width $\sigma$ equals 0.2.

$Wr_i$ is the writhe value of the $i$-th knot. $n_i^{best}$ is the number of the level, $n_i^{best}qW$, closest to $Wr_i$. The summation runs over all alternating prime knots with up to 10 crossings. As seen in Fig.3, the relative average deviation $\langle |dW| \rangle$ of the writhe values from the levels separated by $qW$ displays a clear minimum at $qW$=0.5702, a value very close to the rational $4/7 \cong 0.5714$. At the minimum $\langle |dW| \rangle$=0.075, significantly less than 0.25 expected in absence of the quantization tendency.

Having checked that suggested by Cerf and Stasiak 4/7 writhe quantum produces a writhe levels scheme which fits best the writhe values found by SONO we performed a quantitative analysis of the distribution of observed deviations. Thus, assuming $qW$=4/7 we calculated for each of the analyzed knots the deviation $dW_i$ of its writhe value $Wr_i$ from the closest $n*qW$ writhe level. The plot of the deviations versus the knot number is shown in Fig.4. As seen in the figure, the deviations are spread in an almost uniform manner; their absolute value is never larger than 0.25. The width of the spread is only slightly smaller for smaller knots. To analyze the distribution in a quantitative manner, we divided the [-0.5, 0.5] interval into bins of 0.05 width. Counting knots whose $dW_i$ value were located within consecutive bins and dividing the counts by the total number of analyzed knots we obtained the probability of finding the writhe value within each of the counting bins. The shape of the probability distribution is shown in Fig.5. The half-width $\sigma$ of the Gaussian which fits best the distribution equals 0.203. Moreover, as seen in fig.4, the writhe value of none of the studied knots deviates from the Cerf-Stasiak quantization scheme more than 0.25 $qW$.

## 6. Discussion

Let us summarize results described above.
1. The existence of the writhe quasi-quantization tendency noticed previously within a very limited set of knots [17] has been confirmed within a much broader set of knots: all alternating prime knots with up to 10 crossings.
2. The $qW$=4/7 separation of the writhe levels suggested by Cerf and Stasiak [20] was shown to fit best the writhe data obtained with the use of the SONO algorithm.

The half width $\sigma$ of the distribution of the writhe deviations from the closest Cerf-Stasiak levels was shown to be equal 0.2. In view of the analysis of numerical errors we performed, deviations of such a magnitude cannot be attributed to the inaccuracy of the writhe calculations; they must be seen as the intrinsic property of the most tight conformations found by the SONO algorithm. Will a different algorithm of the determination of the ideal conformations substantially reduce the value of $\sigma$? Will it reduce it to zero? Is the writhe quasi-quantization an approximate or an exact rule? The questions posed above remain open. An independent analysis, performed with the use of a different knot tightening algorithm, could shed more light on them. It seems also that further theoretical considerations along the line presented in ref. 22 should help to understand the origin of the writhe quantization phenomenon. But, whatever the answers to the questions fromulated above, results of the present study confirm beyond any doubt that the writhe of ideal conformations of prime knots shows a strong tendency to group close to well defined equidistant levels.

As mentioned in the introduction, there exist practical implications of the writhe quasi-quantization phenomenon, which were not noticed before. Let us assume that a knot is tied on a rope having a certain internal structure; let for the sake of simplicity it be a bundle of parallel threads. When forming the rope into a knot one wants to perturb as little as possible its internal bundle structure, one should follow the procedure known as the parallel transport. As indicated by Maggs [23], one can show that the writhe of a knot is closely related to the Berry's phase [24].

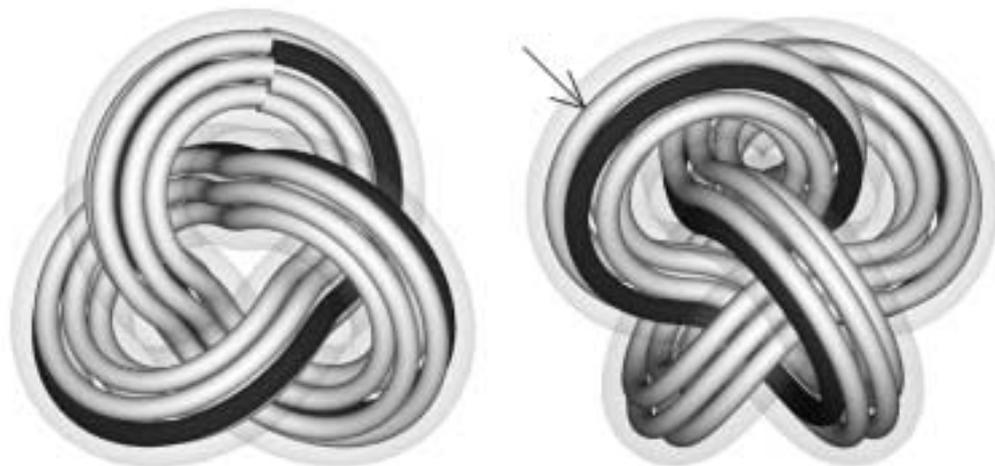

**Fig. 6.** The ideal $3_1$ (left) and $4_6$ (right) knots tied on a rope with an internal structure. The parallel transport of the internal structure of the rope leads in the case of the $3_1$ knot to a distinct misfit of the orientation of the meeting ends. The ends fit perfectly well in the case of the $4_1$ knot – the arrow indicates the hardly visible meeting point.

If connecting the ends of the rope one wants to keep identity of the threads, the angle of the relative orientation of the meeting ends should be equal to a multiple of $2\pi$. This happens when the writhe value of the knot is integer. From such a point of view, the tight conformations of the knots whose writhe values are grouped around a non-integer writhe level are more difficult to tie than the tight conformations of knots whose writhe value stays close to an integer level. Figure 6 presents the ideal $3_1$ and the $4_1$ knots tied on a rope with an internal structure. As seen in the figure, ends of the rope formed using to the parallel transport procedure into the $3_1$ knot do not meet at a proper relative orientation. The knot can be closed only when an additional twist is introduced into the rope. It seems to us that the effect should be taken into consideration in the analysis of knots tied on e.g. the DNA molecules.